\documentclass[twocolumn,prl,showpacs]{revtex4}

\usepackage{graphicx}
\usepackage{rotating}
\usepackage{amsmath}
\usepackage{amsfonts}
\usepackage{amssymb}
\usepackage{enumerate}
\usepackage{longtable}
\usepackage{dcolumn}
\usepackage{bm}

\begin{document}

\title{Pnictides as frustrated quantum antiferromagnet close
to a quantum phase transition}

\author{G\"otz S. Uhrig \footnote{On leave from
Lehrstuhl f\"{u}r Theoretische Physik I,
Technische  Universit\"{a}t Dortmund,
 Otto-Hahn Stra\ss{}e 4, 44221 Dortmund, Germany}}
\email{goetz.uhrig@tu-dortmund.de}
\affiliation{School of Physics, University of New South Wales, 
Kensington 2052, Sydney NSW, Australia}
\author{Michael Holt}
\affiliation{School of Physics, University of New South Wales, 
Kensington 2052, Sydney NSW, Australia}
\author{Jaan Oitmaa}
\affiliation{School of Physics, University of New South Wales, 
Kensington 2052, Sydney NSW, Australia}
\author{Oleg P. Sushkov}
\affiliation{School of Physics, University of New South Wales, 
Kensington 2052, Sydney NSW, Australia}
\author{Rajiv R.P. Singh}
\affiliation{University of California Davis, CA 95616, USA}

\date{\rm\today}

\begin{abstract}
We present a detailed description of the dynamics of the magnetic modes in the
recently discovered superconducting pnictides using 
reliable self-consistent spin-wave theory and series
expansion. Contrary to linear spin-wave theory,
no gapless mode occurs at the N\'eel wave vector.  
We discuss the scenario that the static magnetic moment is
strongly reduced by magnetic fluctuations arising from the 
vicinity to a quantum phase transition. Smoking gun
experiments to verify this scenario are proposed and possible
results are predicted. 
Intriguingly in this scenario, the structural transition at finite temperature 
would be driven by an Ising transition in directional degrees of freedom.
\end{abstract}

\pacs{74.70.-b, 75.10.Jm, 75.40.Gb, 75.30.Ds}


\maketitle

The seminal discovery of superconductivity in the pnictide family
of materials at relatively high temperatures\cite{kamih08} has led to 
tremendous excitement and research activity. This discovery raises many 
fundamental questions. Foremost among them is whether high temperature 
superconductivity in these materials is in some fundamental sense closely 
related to those in the cuprate family of materials. Indeed the quasi-two 
dimensional layered structure for the two families and antiferromagnetism in 
the parent compounds suggests potential similarities. However, many
doubts have also been raised about any correspondence, such as: 
Is the origin and nature of spin fluctuations in the two families related 
given that the parent compounds are metallic in the pnictides whereas they are
insulating in the cuprates? Are the pnictide materials even strongly
correlated or are Local Density Approximation based approaches
adequate? Are spin models appropriate for describing spin-fluctuations
in these materials? In order to address these very basic questions, 
it is necessary to have detailed 
quantitative comparisons between theory and experiments.

\begin{figure}[ht]
    \begin{center}
    \includegraphics[width=0.31\columnwidth,clip]{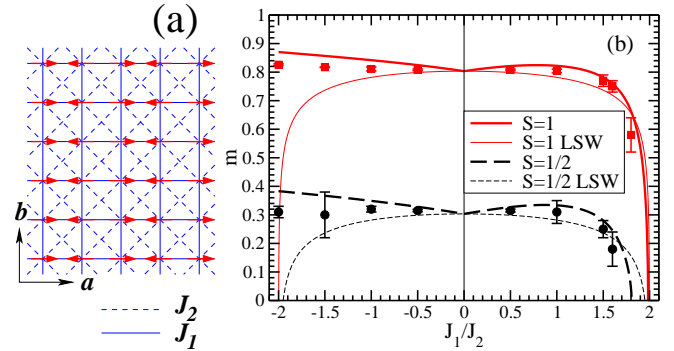}
    \includegraphics[width=0.67\columnwidth,clip]{fig1b}
    \end{center}
    \caption{(color online) Panel (a) considered spin pattern; panel (b) 
      staggered magnetization for $S=1/2$ and $S=1$ as function of the
    ratio of the couplings. For comparison the results of linear spin wave 
    theory (LSW) are included. The self-consistent spin wave theory is
    nicely supported by series expansion about the Ising limit (symbols).}
    \label{fig:magnet}
\end{figure}

On the experimental front,
magnetic long range order was established in LaFeAsO$_{1-x}$F$_x$
by neutron scattering (NS) \cite{cruz08} and by muon spin resonance
($\mu$SR) \cite{klaus08}.
The NS provides evidence for a columnar antiferromagnetic ordering with a 
staggered magnetic moment of $0.36(5)\mu_\text{B}$. For simplicity we
consider here only the square lattice which is formed by the Fe ions
ignoring a small orthorhombic and even monoclinic structural distortion.
Along the a axis the spin directions alternate whereas they are the same
along the b axis, see Fig.\ \ref{fig:magnet}(a). The $\mu$SR also
provides evidence that the spin order is commensurate but with a small
staggered moment of $0.25\mu_\text{B}$.
First results on the dispersion of the magnetic excitations
have just become available \cite{zhao08,ewing08}. 
A tiny anisotropy gap is found to be  $\approx 6$meV, 
the spin wave velocity $v_\perp$ perpendicular to the stripes
to be 205$\pm 20$meV in units of $1/g$, where $g$ is 
the inverse Fe-Fe distance, and a small interplane
coupling $J_z$ is found to be $\approx 5$meV. Results for the parallel
spin wave velocity $v_\parallel$ are not available so far, but they
are expected soon.

Theoretically, the columnar antiferromagnetic ordering was also found 
to be the most stable in band structure calculation \cite{dong08b,yin08}.
So there is agreement on the static structure. But the smallness
of the staggered magnetic moment is a matter of controversy.
On the one hand,
band structure results indicate a local moment of
up to $2.3\mu_\text{B}$ \cite{cao08,ma08a,yin08}. This has led
to the suggestion that the magnetic fluctuations themselves strongly
reduce the static local moment \cite{si08}. We will show that this
scenario implies that the pnictides are in the direct vicinity of
a quantum phase transition. On the other hand,
there are studies suggesting that the strong reduction
of the local magnetic moment can be explained by electronic effects
like hybridization, spin-orbit coupling, and a particular low symmetry
\cite{wu08,yildr08}. Hence, there are two different scenarios:
(i) the local static moment is reduced by the magnetic
fluctuations.  In this case the ratio of couplings must be fixed to an
appropriate value.
(ii) the local electronic orbitals account for the sizable reduction
so that the magnetic couplings are not determined by the value
of the magnetic moment.

Together with upcoming experimental results, our work will 
help decide on the degree of strong correlation and on the
closeness to a quantum phase transition.

The quantitative goal of the present work is three-fold. First, we provide
a theory for the magnetic excitations based on
a minimal spin model, namely the $J_1$-$J_2$ Heisenberg model with
spin $S=1$ ($S=1/2$ results are shown for comparison and to justify
our approximations). Measurements of the spin wave dispersion up to its
maxima will further support the spin-Hamiltonian approach whose validity in
 turn shows that strong correlations dominate the pnictides.
We show that a sizable finite
energy of the spin waves is to be expected at ${\bf q}=(1,1)$
where we denote all wave vectors in units of $\pi/g$. This is in stark
contrast to the results of linear spin-wave theory  (LSW) where
a vanishing spin wave energy is predicted \cite{fang08,yao08}.
Second, we discuss the possibility that the magnetic fluctuations 
reduce the static local moment.
Third, we make quantitative predictions of the dispersion and
of the anisotropy in spin wave velocities along ($v_\parallel$) and across 
($v_\perp$) the magnetic stripes. Measurements of $v_\parallel/v_\perp$ can 
be used to determine the magnetic frustration $J_2/J_1$. 
The spin-wave spectra over the full Brillouin zone (BZ) show clear differences 
between a system deep in the columnar phase and one close to a quantum phase 
transition,  where magnetic fluctuations dramatically reduce the static moment.
This provides a robust experimental way to distinguish the two scenarios (i) 
and (ii).

The parent compound LaFeAsO is not a Mott-Hubbard insulator. 
It is rather a bad metal or semimetal without a Drude peak in the 
conductivity  \cite{dong08b,haule08}.
Even in the magnetically ordered phase the entire Fermi surface
is not gapped \cite{dong08b,ma08a}. Still, in the undoped system the
magnetic excitations are long-lived as they appear as sharp
peaks \cite{zhao08} even at high energies \cite{ewing08}.
So it is justified, though not undisputed \cite{hu08}, 
to start with a model of
well-localized spins coupled by Heisenberg interactions
\begin{equation}
\label{eq:hamilton}
H=J_1\sum_{\langle i,j\rangle} {\bf S}_i\cdot {\bf S}_j
+ J_2\sum_{\langle\langle i,j\rangle\rangle} {\bf S}_i\cdot {\bf S}_j,
\end{equation}
where $\langle i,j\rangle$ stands for nearest neighbors (NN) and
$\langle\langle i,j\rangle\rangle$ for next-nearest neighbors (NNN).
The spin operators read ${\bf S}_i$ and represent $S=1$. This is not completely
obvious in view of the complicated local electronic situations.
But the success of two-band models strongly indicates that $S=1$
is the appropriate choice \cite{raghu08,korsh08}. Furthermore,
the band structure results \cite{cao08,ma08a,yin08}
with a local moment of up to $2.3\mu_\text{B}$
also indicates that there can be up to two electrons aligned.
While our choice \eqref{eq:hamilton} neglects life time effects
due to the decay into particle-hole pairs (Landau damping) 
we expect that the collective magnetic excitations and 
their dispersion are captured.

The choice of a $J_1$-$J_2$ Heisenberg model is justified because the
superexchange is realized mostly via the As ion which sits in the middle
of each Fe plaquette. Hence a NN contribution and a NNN contribution
is to be expected \cite{si08}. Indeed, band structure calculations
show that the NN and the NNN coupling are sizable \cite{yildr08,ma08b,yin08}.
Both $J_1$ and $J_2$ turn out to be antiferromagnetic, i.e., positive
and very similar in value. For this reason, we will choose for 
scenario (ii) the ratio $J_1/J_2=1$.

\begin{figure}[ht]
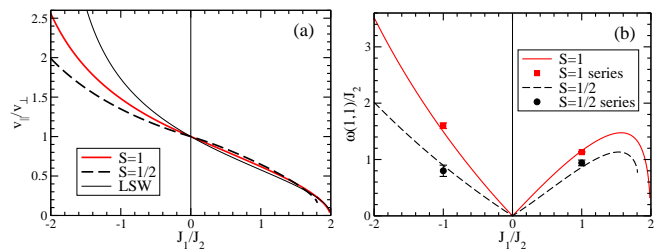

    \begin{center}
     \includegraphics[width=0.50\columnwidth,clip]{fig2a}
     \includegraphics[width=0.48\columnwidth,clip]{fig2b}
    \end{center}
    \caption{(color online) Panel (a) ratio of the spin wave velocities 
      parallel and perpendicular to the spin stripes, i.e., parallel is in 
      $b$ direction and perpendicular is in a direction, see Fig.\ 
      \ref{fig:magnet}(a). We stress that $v_\parallel$, though small, 
      does not vanish where the magnetization vanishes.
      For comparison, LSW data is also included.
      Panel (b) depicts  the spin wave energy $\omega({\bf q})$
      at ${\bf q}=(1,1)$ and ${\bf q}=(0,1)$, respectively, in units of 
      $\pi/g$ ($g$ lattice constant of the assumed square lattice).
      Note that $\omega({\bf q})$ , though small, 
      does not vanish where the magnetization vanishes. The results 
      are nicely corroborated by exemplary series expansion data (symbols).
      }
    \label{fig:ratio}
\end{figure}

The technique employed is self-consistent spin wave theory,
It has been shown previously that this approximation works extremely
well in the columnar, stripe-like phase \cite{oitma96c,singh03}
for $S=1/2$, see also Figs.\ \ref{fig:magnet}(b) and \ref{fig:ratio}(b).
Because spin wave theory can be derived as a
$1/S$ expansion, the results should only improve for
$S=1$.

We have used the Dyson-Maleev as well as the Schwinger 
boson representation \cite{auerb94}. Both yield the same result on the level of
self-consistent mean-field theory. In the symmetry broken
phase the dispersion reads
\begin{subequations}
\label{eq:dispersion}
\begin{eqnarray}
\omega({\bf q}) &=& C J_2\sqrt{A^2-
B^2}\\
A &=& \lambda +x\beta\cos(\pi q_b)\\
B &=& 2\cos(\pi q_a)\cos(\pi q_b)+ x\alpha\cos(\pi q_a)
\label{eq:gapless}
\end{eqnarray}
\end{subequations}
with $x = J_1/J_2$. The expectation values $C$, $\alpha$ and $\beta$ are 
determined from the self-consistency conditions
\begin{subequations}
\begin{eqnarray}
\label{eq:constraint}
2S &=& \partial E_\text{MF}/\partial (CJ_2\lambda)+ 2m\\
2CJ_2 &=& -\partial E_\text{MF}/\partial (CJ_2)+ 4J_2 m\\\
C \alpha J_1 &=& -\partial E_\text{MF}/\partial (C\alpha) +2J_1 m\\
C \beta J_1 &=& \partial E_\text{MF}/\partial (C\beta)+2J_1 m,
\end{eqnarray}
\end{subequations}
where $m$ is the staggered magnetization and the mean-field energy
per spin
\begin{equation}
E_\text{MF} = (2\pi)^{-2}\int_\text{BZ} (\omega({\bf q})-ACJ_2)d^2q
\end{equation}
is used. One integration can be done analytically, the second
numerically. The self-consistency is solved by iteration.
Gaplessness at zero wave vector implies $\lambda =  2+x(\alpha-\beta)$ so that
\eqref{eq:constraint} is used to determine $m$.

The resulting magnetizations are shown in Fig.\ \ref{fig:magnet}(b).
Note the extremely fast vanishing of the magnetization if $x$ approaches
2. The vicinity of the classical first order instability at $x=2$
\cite{weber03} makes $m(x)$
resemble a square root as $x\to x_c$ where $m$ vanishes 
($x_c|_{S=1/2}=1.8057$ and $x_c|_{S=1}=1.9836$) within the
approximation. One may speculate that this is due to the
Ising-type transition related to the breaking of directional
symmetry \cite{chand90,capri04,fang08}, but so far we cannot draw 
a definitive conclusion on this point. But in the light 
of the structural phase transition occurring before 
\cite{cruz08,klaus08} or at \cite{krell08,rotte08}
the magnetic phase transition this aspect is experimentally very interesting.
The structural transition could easily be driven by the Ising transition
in the directional degrees of freedom.

For completeness, we also include results for negative $J_1$.
Around $x\approx -2$ another instability is expected
\cite{shann06}. But interestingly it does not lead to any precursors in
the sublattice magnetization as found from 
self-consistent spin-wave theory and series expansion.
We attribute this behavior to the fact that the classical instability
to the ferromagnetic phase is completely first order in the sense
that there are no precursive fluctuations because the ferromagnetic
phase is free from quantum fluctuations.

The breaking of the directional symmetry implies that the
spin wave velocities depend on direction, see Fig.\ \ref{fig:ratio}(a).
This quantity is a much more robust probe for the 
value of the ratio $x=J_1/J_2$.
The magnetization depends on matrix elements which
in turn can depend on itinerancy, hybridization and other effects.
Energies in contrast only depend on the
Hamiltonian and thus are much less ambiguous.

Furthermore, one notes that \eqref{eq:dispersion}
implies that there is a \emph{finite} excitation energy at
${\bf q}=(\pm 1,\pm 1)$ which is equivalent to ${\bf q}=(0,\pm1)$
if $x(\alpha-\beta) > 0$. The results are plotted in Fig.\ \ref{fig:ratio}(b).
We stress that no dependence of the bare coupling $J_1$
on the bond direction is required.

\begin{figure}[ht]
    \begin{center}
     \includegraphics[width=0.6\columnwidth,clip,angle=270]{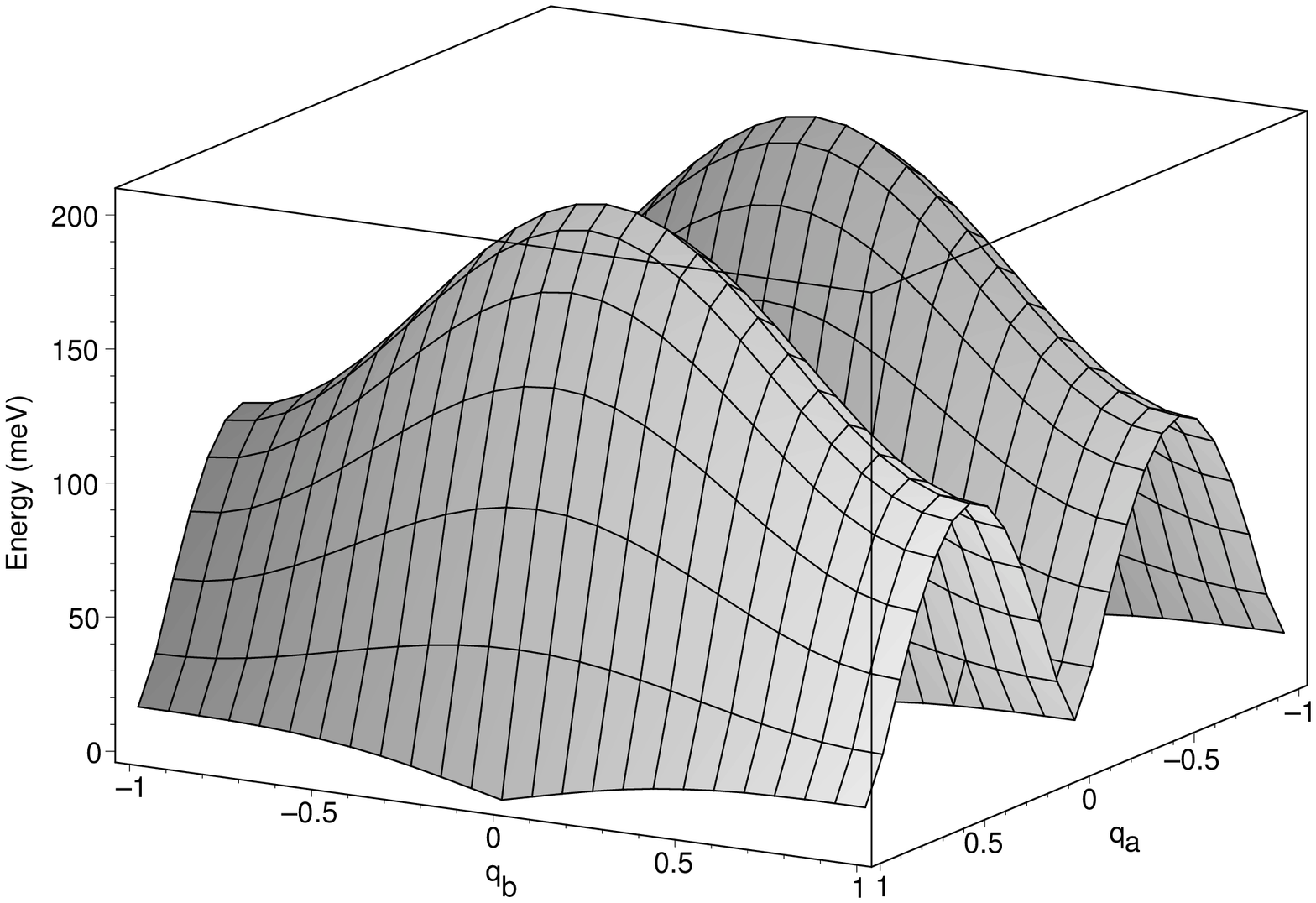}
     \includegraphics[width=0.6\columnwidth,clip,angle=270]{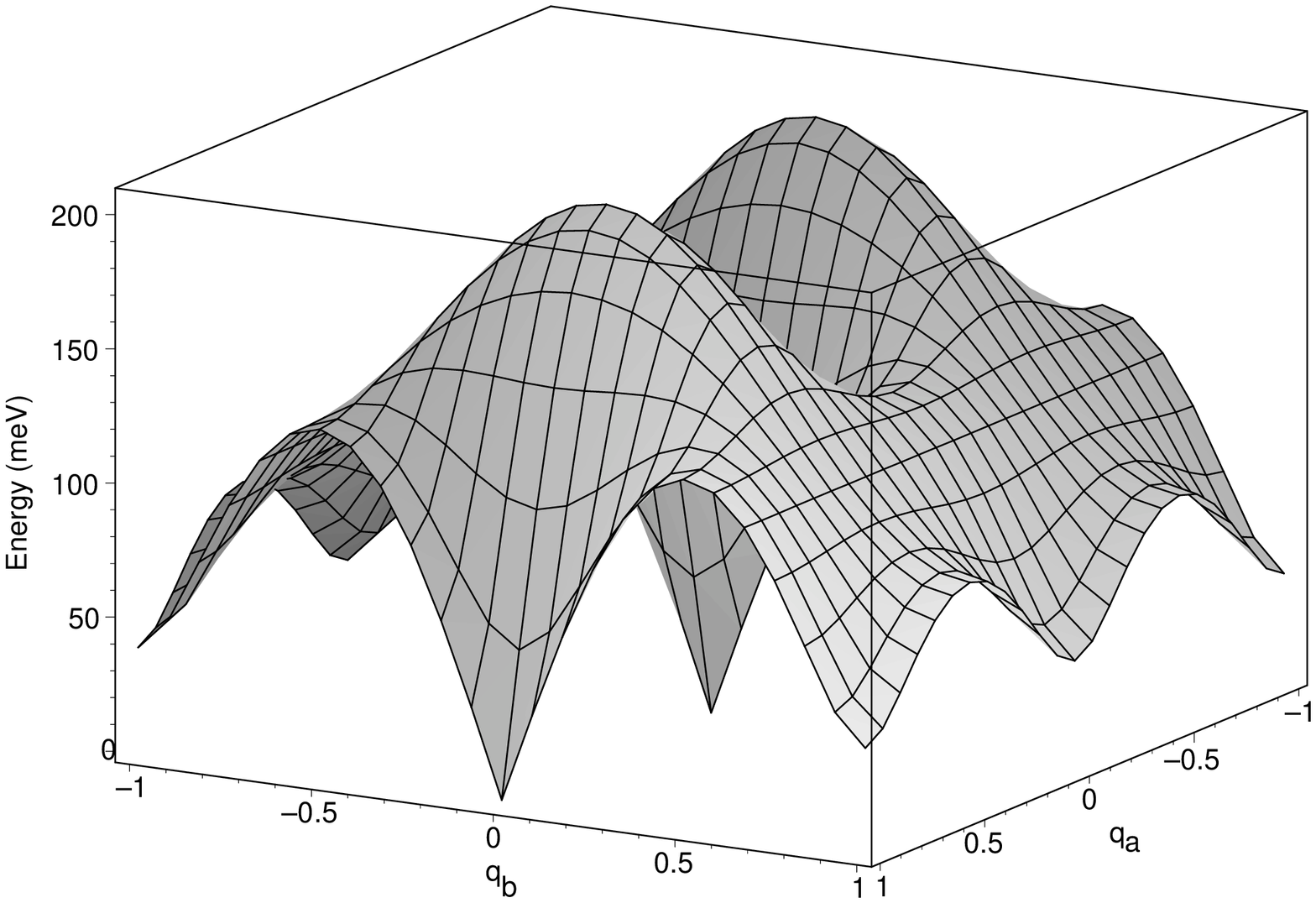}
    \end{center}
    \caption{Upper panel: dispersion of scenario (i) with 
    $S=1$, $J_2=34$meV and $J_1/J_2=1.978$; lower panel:
      dispersion of scenario (ii) with $S=1$, $J_2=33$meV and $J_1/J_2=1$.
      $J_2$ is chosen such that $v_\perp$ equals the experimental value
      \cite{zhao08}.  Wave vectors in units of $\pi/g$.}
    \label{fig:dispersion-3d}
\end{figure}

Now we turn to the two scenarios presented. 
In scenario (i) we attribute
the reduction of $m$ to renormalization by the spin fluctuations.
Equating the ratio $0.36/2.3$ as from experiment \cite{cruz08} and band
structure theory \cite{cao08,ma08a} to $m$ in Fig.\ \ref{fig:magnet}(b)
leads to $J_1=1.978 J_2$. From Fig.\ \ref{fig:magnet}(b) it is obvious
that considerable fine-tuning is needed. The resulting dispersion is
shown in the upper panel in Fig.\ \ref{fig:dispersion-3d}. The lower
panel shows the dispersion for $J_1= J_2$ as suggested by 
band structure calculations \cite{yildr08,ma08b}. In both cases
the overall scale of the coupling is adjusted to fit
to the measured spin velocity $v_\perp$ \cite{zhao08}.
We stress that the value for $J_2$ of about $33$meV agree very well
with the estimates from band structure calculations \cite{yildr08,ma08b}.

\begin{figure}[ht]
    \begin{center}
     \includegraphics[width=0.8\columnwidth,clip]{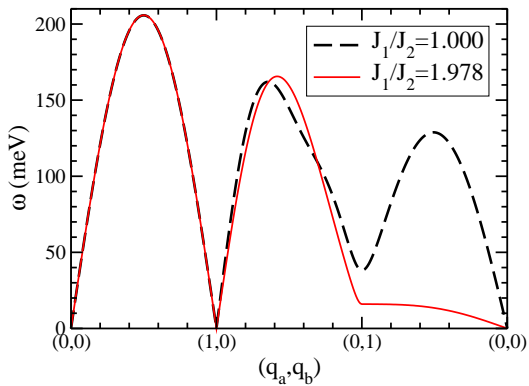}
    \end{center}
    \caption{(color online) Dispersions of both scenarios for
      a generic path through the BZ.  They coincide
    along $(0,0) \to (1,0)$, but differ strongly along $(0,1) \to (0,0)$.
    In LSW the dispersions at $(0,1)$ would vanish spuriously.
    Wave vectors are given in units of $\pi/g$.}
    \label{fig:dispersion-2d}
\end{figure}

Comparing the two panels of Fig.\ \ref{fig:dispersion-3d},
the difference in the dispersion of the spin waves
in both scenarios is striking. Thus a measurement of the 
energetically higher lying modes will easily distinguish
both scenarios. To facilitate the distinction
we plot in Fig.\ \ref{fig:dispersion-2d}
the dispersion along a generic path in the BZ.
The perpendicular spin wave velocity $v_\perp$ is fixed to
its experimental value \cite{zhao08} by the appropriate choice of $J_2$. 
Obviously, the motion across the
stripes (along $a^\star$) is the same for both sets of parameters. The 
important difference occurs in the motion along the stripes 
(along $b^\star$). If the spin fluctuations renormalize
the magnetic moment the spin mode along $b^\star$ is 
extremely soft. Interestingly, this softness 
opens an additional channel for the magnons to decay at
energies of about twice the energy at $(0,1)$, i.e.,
above about 40-50 meV. So in this scenario significant line widths
in inelastic neutron scattering are to be expected.

In conclusion, we presented a quantitative theory for
the dispersion of the spin waves in the recently discovered
superconducting pnictides. It is based on self-consistent
spin wave theory and series expansion for the $S=1$ $J_1$-$J_2$ 
Heisenberg model. Measurements at higher energies will
further support a model of localized spins, for first evidence see Refs.\
 \onlinecite{ni08,ewing08}.  We predict a strong
anisotropy of the spin wave velocities and a finite excitation 
energy at the wave vectors $(0,1)$ and $(1,1)$.

Two scenarios for the strong reduction of the local magnetic
moment are considered. The scenario (i) attributes
the reduction to the magnetic fluctuations. 
We point out that the strongly varying static moments from
$0.25\mu_\text{B}$ \cite{klaus08} over $0.36\mu_\text{B}$ 
\cite{cruz08} to $0.8\mu_\text{B}$  \cite{mcque08}
and $0.9\mu_\text{B}$  \cite{ewing08} finds a natural explanation
if the pnictides are close to the quantum phase transition
at $x\approx 2$ where the renormalized magnetization 
changes very rapidly on small parameter changes.
This scenario implies the fascinating aspect that
the pnictides realize a spin-isotropic system which displays
an Ising transition in the orientation of its ferromagnetic stripes
\cite{chand90,weber03,capri04}.

The alternative scenario (ii) attributes the low magnetic moment to
the local electronic configuration. Then $J_1/J_2\approx 1$
is plausible which does not renormalize the magnetic values
sizably (about 18\%).

The measured anisotropies will allow one to decide how 
close the system is to a quantum phase transition with
fascinating features such as directional Ising transitions. 
A quantitative understanding of spin fluctuations
in the parent materials would help clarify one of the most intriguing
issues in the field, namely, the similarities and differences between
the pnictide and the cuprate family of materials at a fundamental level.

\acknowledgments

We thank Chris  Hamer for very helpful discussions. The financial support (GSU)
by  the Heinrich-Hertz Stiftung NRW and the Gordon 
Godfrey Fund is gratefully acknowledged.


\begin{thebibliography}{30}
\expandafter\ifx\csname natexlab\endcsname\relax\def\natexlab#1{#1}\fi
\expandafter\ifx\csname bibnamefont\endcsname\relax
  \def\bibnamefont#1{#1}\fi
\expandafter\ifx\csname bibfnamefont\endcsname\relax
  \def\bibfnamefont#1{#1}\fi
\expandafter\ifx\csname citenamefont\endcsname\relax
  \def\citenamefont#1{#1}\fi
\expandafter\ifx\csname url\endcsname\relax
  \def\url#1{\texttt{#1}}\fi
\expandafter\ifx\csname urlprefix\endcsname\relax\def\urlprefix{URL }\fi
\providecommand{\bibinfo}[2]{#2}
\providecommand{\eprint}[2][]{\url{#2}}

\bibitem[{\citenamefont{Kamihara et~al.}(2008)\citenamefont{Kamihara, Watanabe,
  Hirano, and Hosono}}]{kamih08}
\bibinfo{author}{\bibfnamefont{Y.}~\bibnamefont{Kamihara}},
  \bibinfo{author}{\bibfnamefont{T.}~\bibnamefont{Watanabe}},
  \bibinfo{author}{\bibfnamefont{M.}~\bibnamefont{Hirano}}, \bibnamefont{and}
  \bibinfo{author}{\bibfnamefont{H.}~\bibnamefont{Hosono}},
  \bibinfo{journal}{J. Am. Chem. Soc.} \textbf{\bibinfo{volume}{130}},
  \bibinfo{pages}{3296} (\bibinfo{year}{2008}).

\bibitem[{\citenamefont{de~la Cruz et~al.}(2008)\citenamefont{de~la Cruz,
  Huang, Lynn, Li, II, Zarestky, Mook, Chen, Luo, Wang et~al.}}]{cruz08}
\bibinfo{author}{\bibfnamefont{C.}~\bibnamefont{de~la Cruz}},
  \bibinfo{author}{\bibfnamefont{Q.}~\bibnamefont{Huang}},
  \bibinfo{author}{\bibfnamefont{J.~W.} \bibnamefont{Lynn}},
  \bibinfo{author}{\bibfnamefont{J.}~\bibnamefont{Li}},
  \bibinfo{author}{\bibfnamefont{W.~R.} \bibnamefont{II}},
  \bibinfo{author}{\bibfnamefont{J.~L.} \bibnamefont{Zarestky}},
  \bibinfo{author}{\bibfnamefont{H.~A.} \bibnamefont{Mook}},
  \bibinfo{author}{\bibfnamefont{G.~F.} \bibnamefont{Chen}},
  \bibinfo{author}{\bibfnamefont{J.~L.} \bibnamefont{Luo}},
  \bibinfo{author}{\bibfnamefont{N.~L.} \bibnamefont{Wang}},
  \bibnamefont{et~al.}, \bibinfo{journal}{Nature}
  \textbf{\bibinfo{volume}{453}}, \bibinfo{pages}{899} (\bibinfo{year}{2008}).

\bibitem[{\citenamefont{Klauss et~al.}(2008)\citenamefont{Klauss, Luekens,
  Klingeler, Hess, Litterst, Kraken, Korshunov, Eremin, Drechsler, Khasanov
  et~al.}}]{klaus08}
\bibinfo{author}{\bibfnamefont{H.-H.} \bibnamefont{Klauss}},
  \bibinfo{author}{\bibfnamefont{H.}~\bibnamefont{Luekens}},
  \bibinfo{author}{\bibfnamefont{R.}~\bibnamefont{Klingeler}},
  \bibinfo{author}{\bibfnamefont{C.}~\bibnamefont{Hess}},
  \bibinfo{author}{\bibfnamefont{F.~J.} \bibnamefont{Litterst}},
  \bibinfo{author}{\bibfnamefont{M.}~\bibnamefont{Kraken}},
  \bibinfo{author}{\bibfnamefont{M.~M.} \bibnamefont{Korshunov}},
  \bibinfo{author}{\bibfnamefont{I.}~\bibnamefont{Eremin}},
  \bibinfo{author}{\bibfnamefont{S.-L.} \bibnamefont{Drechsler}},
  \bibinfo{author}{\bibfnamefont{R.}~\bibnamefont{Khasanov}},
  \bibnamefont{et~al.}, \bibinfo{journal}{Phys. Rev. Lett.}
  \textbf{\bibinfo{volume}{101}}, \bibinfo{pages}{077005}
  (\bibinfo{year}{2008}).

\bibitem[{\citenamefont{Zhao et~al.}(2008)\citenamefont{Zhao, Yao, Li, Hong,
  Chen, Chang, II, Lynn, Mook, Chen et~al.}}]{zhao08}
\bibinfo{author}{\bibfnamefont{J.}~\bibnamefont{Zhao}},
  \bibinfo{author}{\bibfnamefont{D.-X.} \bibnamefont{Yao}},
  \bibinfo{author}{\bibfnamefont{S.}~\bibnamefont{Li}},
  \bibinfo{author}{\bibfnamefont{T.}~\bibnamefont{Hong}},
  \bibinfo{author}{\bibfnamefont{Y.}~\bibnamefont{Chen}},
  \bibinfo{author}{\bibfnamefont{S.}~\bibnamefont{Chang}},
  \bibinfo{author}{\bibfnamefont{W.~R.} \bibnamefont{II}},
  \bibinfo{author}{\bibfnamefont{J.~W.} \bibnamefont{Lynn}},
  \bibinfo{author}{\bibfnamefont{H.~A.} \bibnamefont{Mook}},
  \bibinfo{author}{\bibfnamefont{G.~F.} \bibnamefont{Chen}},
  \bibnamefont{et~al.}, \bibinfo{journal}{Phys. Rev. Lett.}
  \textbf{\bibinfo{volume}{101}}, \bibinfo{pages}{167203}
  (\bibinfo{year}{2008}).

\bibitem[{\citenamefont{Ewings et~al.}(2008)\citenamefont{Ewings, Perring,
  Bewley, Guidi, Pitcher, Parker, Clarke, and Boothroyd}}]{ewing08}
\bibinfo{author}{\bibfnamefont{R.~A.} \bibnamefont{Ewings}},
  \bibinfo{author}{\bibfnamefont{T.~G.} \bibnamefont{Perring}},
  \bibinfo{author}{\bibfnamefont{R.~I.} \bibnamefont{Bewley}},
  \bibinfo{author}{\bibfnamefont{T.}~\bibnamefont{Guidi}},
  \bibinfo{author}{\bibfnamefont{M.~J.} \bibnamefont{Pitcher}},
  \bibinfo{author}{\bibfnamefont{D.~R.} \bibnamefont{Parker}},
  \bibinfo{author}{\bibfnamefont{S.~J.} \bibnamefont{Clarke}},
  \bibnamefont{and} \bibinfo{author}{\bibfnamefont{A.~T.}
  \bibnamefont{Boothroyd}}, \bibinfo{volume}{arXiv:0808.2836}.

\bibitem[{\citenamefont{Dong et~al.}(2008)\citenamefont{Dong, Zhand, Xu, Li,
  Li, Hu, Wu, Chen, Dai, Luo et~al.}}]{dong08b}
\bibinfo{author}{\bibfnamefont{J.}~\bibnamefont{Dong}},
  \bibinfo{author}{\bibfnamefont{H.~J.} \bibnamefont{Zhand}},
  \bibinfo{author}{\bibfnamefont{G.}~\bibnamefont{Xu}},
  \bibinfo{author}{\bibfnamefont{Z.}~\bibnamefont{Li}},
  \bibinfo{author}{\bibfnamefont{G.}~\bibnamefont{Li}},
  \bibinfo{author}{\bibfnamefont{W.~Z.} \bibnamefont{Hu}},
  \bibinfo{author}{\bibfnamefont{D.}~\bibnamefont{Wu}},
  \bibinfo{author}{\bibfnamefont{G.~F.} \bibnamefont{Chen}},
  \bibinfo{author}{\bibfnamefont{X.}~\bibnamefont{Dai}},
  \bibinfo{author}{\bibfnamefont{J.~L.} \bibnamefont{Luo}},
  \bibnamefont{et~al.}, \bibinfo{journal}{Europhys. Lett.}
  \textbf{\bibinfo{volume}{83}}, \bibinfo{pages}{27006} (\bibinfo{year}{2008}).

\bibitem[{\citenamefont{Z.P.Yin et~al.}(2008)\citenamefont{Z.P.Yin, Leb\`egue,
  Han, Neal, Savrasov, and Pickett}}]{yin08}
\bibinfo{author}{\bibnamefont{Z.P.Yin}},
  \bibinfo{author}{\bibfnamefont{S.}~\bibnamefont{Leb\`egue}},
  \bibinfo{author}{\bibfnamefont{M.~J.} \bibnamefont{Han}},
  \bibinfo{author}{\bibfnamefont{B.~P.} \bibnamefont{Neal}},
  \bibinfo{author}{\bibfnamefont{S.~Y.} \bibnamefont{Savrasov}},
  \bibnamefont{and} \bibinfo{author}{\bibfnamefont{W.~E.}
  \bibnamefont{Pickett}}, \bibinfo{journal}{Phys. Rev. Lett.}
  \textbf{\bibinfo{volume}{101}}, \bibinfo{pages}{047001}
  (\bibinfo{year}{2008}).

\bibitem[{\citenamefont{Cao et~al.}(2008)\citenamefont{Cao, Hirschfeld, and
  Cheng}}]{cao08}
\bibinfo{author}{\bibfnamefont{C.}~\bibnamefont{Cao}},
  \bibinfo{author}{\bibfnamefont{P.~J.} \bibnamefont{Hirschfeld}},
  \bibnamefont{and} \bibinfo{author}{\bibfnamefont{H.-P.} \bibnamefont{Cheng}},
  \bibinfo{journal}{Phys. Rev. B} \textbf{\bibinfo{volume}{77}},
  \bibinfo{pages}{220506} (\bibinfo{year}{2008}).

\bibitem[{\citenamefont{Ma and Lu}(2008)}]{ma08a}
\bibinfo{author}{\bibfnamefont{F.}~\bibnamefont{Ma}} \bibnamefont{and}
  \bibinfo{author}{\bibfnamefont{Z.-Y.} \bibnamefont{Lu}},
  \bibinfo{journal}{Phys. Rev. B} \textbf{\bibinfo{volume}{78}},
  \bibinfo{pages}{033111} (\bibinfo{year}{2008}).

\bibitem[{\citenamefont{Si and Abrahams}(2008)}]{si08}
\bibinfo{author}{\bibfnamefont{Q.}~\bibnamefont{Si}} \bibnamefont{and}
  \bibinfo{author}{\bibfnamefont{E.}~\bibnamefont{Abrahams}},
  \bibinfo{journal}{Phys. Rev. Lett.} \textbf{\bibinfo{volume}{101}},
  \bibinfo{pages}{076401} (\bibinfo{year}{2008}).

\bibitem[{\citenamefont{Wu et~al.}(2008)\citenamefont{Wu, Phillips, and
  Neto}}]{wu08}
\bibinfo{author}{\bibfnamefont{J.}~\bibnamefont{Wu}},
  \bibinfo{author}{\bibfnamefont{P.}~\bibnamefont{Phillips}}, \bibnamefont{and}
  \bibinfo{author}{\bibfnamefont{A.~H.~C.} \bibnamefont{Neto}},
  \bibinfo{journal}{Phys. Rev. Lett.} \textbf{\bibinfo{volume}{101}},
  \bibinfo{pages}{126401} (\bibinfo{year}{2008}).

\bibitem[{\citenamefont{Yildrim}(2008)}]{yildr08}
\bibinfo{author}{\bibfnamefont{T.}~\bibnamefont{Yildrim}},
  \bibinfo{journal}{Phys. Rev. Lett.} \textbf{\bibinfo{volume}{101}},
  \bibinfo{pages}{057010} (\bibinfo{year}{2008}).

\bibitem[{\citenamefont{Fang et~al.}(2008)\citenamefont{Fang, Yao, Tsai, Hu,
  and Kivelson}}]{fang08}
\bibinfo{author}{\bibfnamefont{C.}~\bibnamefont{Fang}},
  \bibinfo{author}{\bibfnamefont{H.}~\bibnamefont{Yao}},
  \bibinfo{author}{\bibfnamefont{W.}~\bibnamefont{Tsai}},
  \bibinfo{author}{\bibfnamefont{J.}~\bibnamefont{Hu}}, \bibnamefont{and}
  \bibinfo{author}{\bibfnamefont{S.~A.} \bibnamefont{Kivelson}},
  \bibinfo{journal}{Phys. Rev. B} \textbf{\bibinfo{volume}{77}},
  \bibinfo{pages}{224509} (\bibinfo{year}{2008}).

\bibitem[{\citenamefont{Yao and Carlson}(2008)}]{yao08}
\bibinfo{author}{\bibfnamefont{D.-X.} \bibnamefont{Yao}} \bibnamefont{and}
  \bibinfo{author}{\bibfnamefont{E.~W.} \bibnamefont{Carlson}},
  \bibinfo{journal}{Phys. Rev. B} \textbf{\bibinfo{volume}{78}},
  \bibinfo{pages}{052507} (\bibinfo{year}{2008}).

\bibitem[{\citenamefont{Haule et~al.}(2008)\citenamefont{Haule, Shim, and
  Kotliar}}]{haule08}
\bibinfo{author}{\bibfnamefont{K.}~\bibnamefont{Haule}},
  \bibinfo{author}{\bibfnamefont{J.~H.} \bibnamefont{Shim}}, \bibnamefont{and}
  \bibinfo{author}{\bibfnamefont{G.}~\bibnamefont{Kotliar}},
  \bibinfo{journal}{Phys. Rev. Lett.} \textbf{\bibinfo{volume}{100}},
  \bibinfo{pages}{226402} (\bibinfo{year}{2008}).

\bibitem[{\citenamefont{Hu et~al.}(2008)\citenamefont{Hu, Dong, Li, Li, Zheng,
  Chen, Luo, and Wang}}]{hu08}
\bibinfo{author}{\bibfnamefont{W.~Z.} \bibnamefont{Hu}},
  \bibinfo{author}{\bibfnamefont{J.}~\bibnamefont{Dong}},
  \bibinfo{author}{\bibfnamefont{G.}~\bibnamefont{Li}},
  \bibinfo{author}{\bibfnamefont{Z.}~\bibnamefont{Li}},
  \bibinfo{author}{\bibfnamefont{P.}~\bibnamefont{Zheng}},
  \bibinfo{author}{\bibfnamefont{G.~F.} \bibnamefont{Chen}},
  \bibinfo{author}{\bibfnamefont{J.~L.} \bibnamefont{Luo}}, \bibnamefont{and}
  \bibinfo{author}{\bibfnamefont{N.~L.} \bibnamefont{Wang}},
  \bibinfo{pages}{arXiv:0806.2652}.

\bibitem[{\citenamefont{Raghu et~al.}(2008)\citenamefont{Raghu, Qi, Liu,
  Scalapino, and Zhang}}]{raghu08}
\bibinfo{author}{\bibfnamefont{S.}~\bibnamefont{Raghu}},
  \bibinfo{author}{\bibfnamefont{X.-L.} \bibnamefont{Qi}},
  \bibinfo{author}{\bibfnamefont{C.-X.} \bibnamefont{Liu}},
  \bibinfo{author}{\bibfnamefont{D.~C.} \bibnamefont{Scalapino}},
  \bibnamefont{and} \bibinfo{author}{\bibfnamefont{S.-C.} \bibnamefont{Zhang}},
  \bibinfo{journal}{Phys. Rev. B} \textbf{\bibinfo{volume}{77}},
  \bibinfo{pages}{220503(R)} (\bibinfo{year}{2008}).

\bibitem[{\citenamefont{Korshunov and Eremin}(2008)}]{korsh08}
\bibinfo{author}{\bibfnamefont{M.~M.} \bibnamefont{Korshunov}}
  \bibnamefont{and} \bibinfo{author}{\bibfnamefont{I.}~\bibnamefont{Eremin}},
  \bibinfo{journal}{arXiv:0804.1793}.

\bibitem[{\citenamefont{Ma et~al.}(2008)\citenamefont{Ma, Lu, and
  Xiang}}]{ma08b}
\bibinfo{author}{\bibfnamefont{F.}~\bibnamefont{Ma}},
  \bibinfo{author}{\bibfnamefont{Z.-Y.} \bibnamefont{Lu}}, \bibnamefont{and}
  \bibinfo{author}{\bibfnamefont{T.}~\bibnamefont{Xiang}},
  \bibinfo{journal}{arXiv:0804.3370}.

\bibitem[{\citenamefont{Oitmaa and Weihong}(1996)}]{oitma96c}
\bibinfo{author}{\bibfnamefont{J.}~\bibnamefont{Oitmaa}} \bibnamefont{and}
  \bibinfo{author}{\bibfnamefont{Z.}~\bibnamefont{Weihong}},
  \bibinfo{journal}{Phys. Rev. B} \textbf{\bibinfo{volume}{54}},
  \bibinfo{pages}{3022} (\bibinfo{year}{1996}).

\bibitem[{\citenamefont{Singh et~al.}(2003)\citenamefont{Singh, Zheng, Oitmaa,
  Sushkov, and Hamer}}]{singh03}
\bibinfo{author}{\bibfnamefont{R.~R.~P.} \bibnamefont{Singh}},
  \bibinfo{author}{\bibfnamefont{W.}~\bibnamefont{Zheng}},
  \bibinfo{author}{\bibfnamefont{J.}~\bibnamefont{Oitmaa}},
  \bibinfo{author}{\bibfnamefont{O.~P.} \bibnamefont{Sushkov}},
  \bibnamefont{and} \bibinfo{author}{\bibfnamefont{C.~J.} \bibnamefont{Hamer}},
  \bibinfo{journal}{Phys. Rev. Lett.} \textbf{\bibinfo{volume}{91}},
  \bibinfo{pages}{017201} (\bibinfo{year}{2003}).

\bibitem[{\citenamefont{Auerbach}(1994)}]{auerb94}
\bibinfo{author}{\bibfnamefont{A.}~\bibnamefont{Auerbach}},
  \emph{\bibinfo{title}{Interacting Electrons and Quantum Magnetism}}, Graduate
  Texts in Contemporary Physics (\bibinfo{publisher}{Springer},
  \bibinfo{address}{New York}, \bibinfo{year}{1994}).

\bibitem[{\citenamefont{Weber et~al.}(2003)\citenamefont{Weber, Capriotti,
  Misguich, Becca, Elhajal, and Mila}}]{weber03}
\bibinfo{author}{\bibfnamefont{C.}~\bibnamefont{Weber}},
  \bibinfo{author}{\bibfnamefont{L.}~\bibnamefont{Capriotti}},
  \bibinfo{author}{\bibfnamefont{G.}~\bibnamefont{Misguich}},
  \bibinfo{author}{\bibfnamefont{F.}~\bibnamefont{Becca}},
  \bibinfo{author}{\bibfnamefont{M.}~\bibnamefont{Elhajal}}, \bibnamefont{and}
  \bibinfo{author}{\bibfnamefont{F.}~\bibnamefont{Mila}},
  \bibinfo{journal}{Phys. Rev. Lett.} \textbf{\bibinfo{volume}{91}},
  \bibinfo{pages}{177202} (\bibinfo{year}{2003}).

\bibitem[{\citenamefont{Chandra et~al.}(1990)\citenamefont{Chandra, Coleman,
  and Larkin}}]{chand90}
\bibinfo{author}{\bibfnamefont{P.}~\bibnamefont{Chandra}},
  \bibinfo{author}{\bibfnamefont{P.}~\bibnamefont{Coleman}}, \bibnamefont{and}
  \bibinfo{author}{\bibfnamefont{A.~I.} \bibnamefont{Larkin}},
  \bibinfo{journal}{Phys. Rev. Lett.} \textbf{\bibinfo{volume}{64}},
  \bibinfo{pages}{88} (\bibinfo{year}{1990}).

\bibitem[{\citenamefont{Capriotti et~al.}(2004)\citenamefont{Capriotti, Fubini,
  Roscilde, and Tognetti}}]{capri04}
\bibinfo{author}{\bibfnamefont{L.}~\bibnamefont{Capriotti}},
  \bibinfo{author}{\bibfnamefont{A.}~\bibnamefont{Fubini}},
  \bibinfo{author}{\bibfnamefont{T.}~\bibnamefont{Roscilde}}, \bibnamefont{and}
  \bibinfo{author}{\bibfnamefont{V.}~\bibnamefont{Tognetti}},
  \bibinfo{journal}{Phys. Rev. Lett.} \textbf{\bibinfo{volume}{92}},
  \bibinfo{pages}{157202} (\bibinfo{year}{2004}).

\bibitem[{\citenamefont{Krellner et~al.}(2008)\citenamefont{Krellner,
  Caroca-Canales, Jesche, Rosner, Ormeci, and Geibel}}]{krell08}
\bibinfo{author}{\bibfnamefont{C.}~\bibnamefont{Krellner}},
  \bibinfo{author}{\bibfnamefont{N.}~\bibnamefont{Caroca-Canales}},
  \bibinfo{author}{\bibfnamefont{A.}~\bibnamefont{Jesche}},
  \bibinfo{author}{\bibfnamefont{H.}~\bibnamefont{Rosner}},
  \bibinfo{author}{\bibfnamefont{A.}~\bibnamefont{Ormeci}}, \bibnamefont{and}
  \bibinfo{author}{\bibfnamefont{C.}~\bibnamefont{Geibel}},
  \bibinfo{journal}{Phys. Rev. B} \textbf{\bibinfo{volume}{78}},
  \bibinfo{pages}{100504} (\bibinfo{year}{2008}).

\bibitem[{\citenamefont{Rotter et~al.}(2008)\citenamefont{Rotter, Tegel,
  Johrendt, Schellenberg, Hermes, and Pöttgen}}]{rotte08}
\bibinfo{author}{\bibfnamefont{M.}~\bibnamefont{Rotter}},
  \bibinfo{author}{\bibfnamefont{M.}~\bibnamefont{Tegel}},
  \bibinfo{author}{\bibfnamefont{D.}~\bibnamefont{Johrendt}},
  \bibinfo{author}{\bibfnamefont{I.}~\bibnamefont{Schellenberg}},
  \bibinfo{author}{\bibfnamefont{W.}~\bibnamefont{Hermes}}, \bibnamefont{and}
  \bibinfo{author}{\bibfnamefont{R.}~\bibnamefont{Pöttgen}},
  \bibinfo{journal}{Phys. Rev. B} \textbf{\bibinfo{volume}{78}},
  \bibinfo{pages}{020503(R)} (\bibinfo{year}{2008}).

\bibitem[{\citenamefont{Shannon et~al.}(2006)\citenamefont{Shannon, Momoi, and
  Sindzingre}}]{shann06}
\bibinfo{author}{\bibfnamefont{N.}~\bibnamefont{Shannon}},
  \bibinfo{author}{\bibfnamefont{T.}~\bibnamefont{Momoi}}, \bibnamefont{and}
  \bibinfo{author}{\bibfnamefont{P.}~\bibnamefont{Sindzingre}},
  \bibinfo{journal}{Phys. Rev. Lett.} \textbf{\bibinfo{volume}{96}},
  \bibinfo{pages}{027213} (\bibinfo{year}{2006}).

\bibitem[{\citenamefont{Ni et~al.}(1971)\citenamefont{Ni, Bud'ko, Kreyssig,
  Nandi, Rustan, Goldman, Gupta, Corbett, Kracher, and Canfield}}]{ni08}
\bibinfo{author}{\bibfnamefont{N.}~\bibnamefont{Ni}},
  \bibinfo{author}{\bibfnamefont{S.~L.} \bibnamefont{Bud'ko}},
  \bibinfo{author}{\bibfnamefont{A.}~\bibnamefont{Kreyssig}},
  \bibinfo{author}{\bibfnamefont{S.}~\bibnamefont{Nandi}},
  \bibinfo{author}{\bibfnamefont{G.~E.} \bibnamefont{Rustan}},
  \bibinfo{author}{\bibfnamefont{A.~I.} \bibnamefont{Goldman}},
  \bibinfo{author}{\bibfnamefont{S.}~\bibnamefont{Gupta}},
  \bibinfo{author}{\bibfnamefont{J.~D.} \bibnamefont{Corbett}},
  \bibinfo{author}{\bibfnamefont{A.}~\bibnamefont{Kracher}}, \bibnamefont{and}
  \bibinfo{author}{\bibfnamefont{P.~C.} \bibnamefont{Canfield}},
  \bibinfo{journal}{Phys. Rev. B} \textbf{\bibinfo{volume}{78}},
  \bibinfo{pages}{014507} (\bibinfo{year}{1971}).

\bibitem[{\citenamefont{McQueeney et~al.}(2008)\citenamefont{McQueeney, Diallo,
  Antropov, Samolyuk, Broholm, Ni, Nandi, Yethiraj, Zarestky, Pulikkotil
  et~al.}}]{mcque08}
\bibinfo{author}{\bibfnamefont{R.~J.} \bibnamefont{McQueeney}},
  \bibinfo{author}{\bibfnamefont{S.~O.} \bibnamefont{Diallo}},
  \bibinfo{author}{\bibfnamefont{V.~P.} \bibnamefont{Antropov}},
  \bibinfo{author}{\bibfnamefont{G.}~\bibnamefont{Samolyuk}},
  \bibinfo{author}{\bibfnamefont{C.}~\bibnamefont{Broholm}},
  \bibinfo{author}{\bibfnamefont{N.}~\bibnamefont{Ni}},
  \bibinfo{author}{\bibfnamefont{S.}~\bibnamefont{Nandi}},
  \bibinfo{author}{\bibfnamefont{M.}~\bibnamefont{Yethiraj}},
  \bibinfo{author}{\bibfnamefont{J.~L.} \bibnamefont{Zarestky}},
  \bibinfo{author}{\bibfnamefont{J.~J.} \bibnamefont{Pulikkotil}},
  \bibnamefont{et~al.}, {\bibinfo{volume}{arXiv:0809.1410}}.

\end{thebibliography}

\end{document}